\documentclass{article}

\usepackage{arxiv}

\usepackage[utf8]{inputenc} 
\usepackage[T1]{fontenc}    
\usepackage{hyperref}       
\usepackage{url}            
\usepackage{booktabs}       
\usepackage{amsfonts}       
\usepackage{nicefrac}       
\usepackage{microtype}      
\usepackage{lipsum}
\usepackage{graphicx}
\usepackage{circledtext}
\graphicspath{ {./images/} }

\title{Mathematical representation of bias and nudges centered on intangible goods using quantum information theory}

\author{
 Misao Fukuda \\
   \\
  \texttt{fdk.sm01@gmail.com} 
}
\usepackage{amsmath}
\begin{document}
\maketitle
\begin{abstract}
This study aims to explore whether the relationship between bias and nudges can be mathematically expressed in terms of quantum information theory, particularly by means of individually customized nudges. Based on the value function of customer satisfaction, which is subject to uncertainty due to its subjective nature, a model of bias and nudges is proposed that considers the environment for intangible goods. Then, by defining an index of nudges from the mathematical properties of the value function obtained from this economic model, a model can be expressed that has a mathematical structure of the same nature as nudges expressed in standard economics, where welfare is impaired by bias from the mathematical structure of the gross social surplus derived as a social welfare function. Moreover, the mathematical structure of the gross social surplus can be made larger than that in standard economics, adding knowledge about the mathematical design of nudges as individually customized customer experiences. Considering information technology within the context of business intelligence derived from service science and service engineering, the connection between relevant variables and empirical validation is explained. This increases the feasibility of the economic model based on quantum information theory and the mathematical design of customized nudges. 
\end{abstract}


\section{Introduction}
 Based on quantum information theory, this study examines the subjective level of customer satisfaction with intangible goods intended to provide positive customer experiences and thereby suggests solutions to issues regarding the possibility of constructing a mathematical economic model of bias and nudges and its application to intangible goods, which account for a growing share of the economy.
 
According to Thaler and Sunstein (2008), a nudge is an intervention “that alters people’s behavior in a predictable way without forbidding any options or significantly changing their economic incentives.” In the framework of nudges by Jimenez-Gomez (2018), individuals experience internalities (costs that might be imposed on the future self), and their choices and welfare depend on the environment, which nudges can alter. He showed that individual preferences do not need to be fully recovered in social planning, which would be difficult to do because of internalities. In heterogeneous populations, the optimal nudge balances the tradeoff between correcting the internalities of biased individuals and the psychological costs of nudges to all individuals (Jimenez-Gomez, 2018). Nudges have been applied in various domains, including nutrition (List and Samek, 2015), retirement savings (Bernheim et al., 2011), and energy conservation (Alcott and Taubinsky, 2015). Although nudges have been studied and implemented in practical settings, theorical study of nudges by formal economic analysis remains insufficient. One challenge is that formal analyses typically assume infallible and entirely rational agents, making nudges unnecessary. However, if agents are assumed to be fallible, it becomes possible to analyze settings where nudges can increase welfare by utilizing the concepts of experienced utility and decision utility.

The present study uses a standard economics framework to define the difference between decision utility and experienced utility as an internality and mathematically analyzes nudges under several assumptions. The result suggests that for optimal nudges that improve welfare, there is a trade-off between correcting biased individuals’ internalities and the psychological costs of nudges to all individuals, as suggested by Jimenez-Gomez (2018).

There is skepticism about nudges, and reasons why their external validity has not been established include that nudges are context-dependent, and that bottlenecks vary across individuals due to their heterogeneity. In social experiments on nudges, evaluations of effectiveness are needed that include measurement of the quantity or proportion of people who took certain actions due to the interventions (Kawasaki and Tadano, 2023). 

Quantum information theory, which lays the foundation for quantum computers, is based on the problem that observation of an object changes its state (Nielsen and Chuang, 2000). Quantum decision theory is an application of quantum information theory to social science. A mathematical expression of a person’s irrational decisions has been proposed by Tversky and Kahneman (1992) and studied in the field of behavioral economics (Cheon and Takahashi, 2010; Fukuda, 2022, 2023, 2024; Takahashi, 2013; Yukalov and Sornette, 2017). Fukuda (2022) set up and analyzed an economic model of intangible goods based on the theory of observation of quantum information. That study derived the value function in prospect theory and found its correspondence to principles of behavioral economics, which violates the independence axiom. Thus, it is now increasingly plausible that this approach based on quantum information can potentially be used as an economic model for the theory of designing intangible goods and as a mathematical model for designing customer experiences. 

However, to the author’s knowledge, no economic analyses of bias and nudges have been carried out that specify the rate of behavioral improvement with a mathematical behavioral economic model of nudges that draws on quantum information theory and enables mathematical representation of context and heterogeneity.

In terms of research methodology, this study follows previous work that drew on quantum information theory to quantify customer satisfaction. Here the relationship between bias and nudges are explored and the possibility of mathematically representing individually customized nudges is examined. 

In light of the unanswered questions in previous studies, this study treats the subjective experience of intangible goods as a rotation in the Hilbert space, which can mathematically represent context and heterogeneity. In doing this, the aim is to provide new insights into the mathematical structure of nudges by specifying the behavioral improvement rate based on a mathematical model of bias and nudges associated with intangible goods that draws on quantum information theory.

Given that the share of intangible goods in the economy has rapidly increased in recent years, there is now an urgent need to develop scientific approaches to intangible goods rather than relying on experience and intuition, and efforts such as this study have become increasingly necessary.

The findings of this study are as follows. For intangible goods, when the relevant environment is considered, the peak-end rule from behavioral economics emerges when customers’ perception of intangible goods is mathematically entangled with their perception of the relevant environment. The repeat rate is instrumental in understanding bias and nudges, and can serve as an indicator of the rate of behavioral improvement due to nudges. The model analyzed in this study and nudges considered in standard economics share a common mathematical structure. Welfare loss due to bias can be expressed based on the mathematical structure of gross social surplus derived as the social welfare function. Gross social surplus can be larger here than in the case of standard economics. Designing customized customer experiences is possible by mathematically expressing a default effect of nudges. Considering information technology (IT) within the context of business intelligence derived from service science and service engineering, this paper explains the connection between relevant variables and empirical validation.

The new findings of this study regarding the mathematical relationships of nudges are based on the relationship between bias and nudges linked to intangible goods (which, in turn, is based on quantum information theory and reflects the relevant environment) and can increase the feasibility of mathematical design of customer experiences by enabling prediction of the rate of behavioral improvement due to nudges. This may contribute to further development of today’s economy, which has seen an increasing share of intangible goods.

\section{Mathematics of Intangible Goods}
\label{sec:headings}

\subsection{Definitions of Terms and Variables in This Study}

This section defines the variables used in this paper and in Fukuda (2022, 2023, 2024) and discusses their empirical validation. First, the variables are defined below.

$|b_i|^2$, $|b_{ri}|^2$: Magnitude of the impact of process $i$ on the intangible good, as perceived by the customer before experiencing (observing) the good

$|b_{rei}|^2$: Magnitude of the impact of process $i$ on the environment (rules), as perceived by the customer prior to the experience (observation)

$|\alpha_i|^2$, $|\alpha_{ri}|^2$: Magnitude of the impact of process $i$ on the intangible good, as perceived by the customer following the experience (observation)

$|\alpha_{rei}|^2$: Magnitude of the impact of process $i$ on the environment (rules), as perceived by the customer following the experience (observation)

$\theta_i$, $\theta_{ri}$: Magnitude of prior expectations associated with process $i$ regarding the intangible good, as perceived by the customer prior to the experience (observation)

$\theta_{rei}$: Magnitude of prior expectations associated with process $i$ regarding the environment (rules), as perceived by the customer prior to the experience (observation)

$\phi_i$, $\phi_{ri}$: Magnitude of the reputation associated with process $i$, as perceived by the customer following the experience (observation).

$\phi_{rei}$: Magnitude of the reputation associated with process $i$ regarding the environment (rules), as perceived by the customer following the experience (observation)

$\theta_i-\phi_i$: Level of product quality associated with process $i$, as perceived by the customer following the experience (observation)

$n$: Number of repeats

$N$: Number of processes

$t$: Scaling factor for normalizing the sum of the values 
$|\alpha_i|^2$ across all processes $i$ to a maximum of $N$, while preserving the proportions of the values

$p_{n}{(\uparrow)}$: Weight for satisfaction at the $n$th repeat of the intangible good

$p_{n}{(\downarrow)}$: Weight for dissatisfaction at the $n$th repeat of the intangible good

$p_{n+1|n}(\uparrow|\uparrow)$: Weight for satisfaction at the ($n+1$)th repeat, given that the $n$th repeat resulted in satisfaction

$p_{n+1|n}(\uparrow|\downarrow)$: Weight for satisfaction at the ($n+1$)th repeat, given that the $n$th repeat resulted in dissatisfaction

$p_{r}{(\downarrow)}$: Repeat rate, or more generally, the rate of behavioral change, as perceived by the customer following the experience (observation)

$p_{r}{(\uparrow)}$: The churn rate (particularly in the case of subscription services), as perceived by the customer following the experience (observation)

$q_{n+1}$: Correction term representing emotional satisfaction at the $(n+1)$th repeat

$S_{merit}$, $S_{merit}^{R}$: Magnitude of customer satisfaction, as perceived by the customer following the experience (observation)

$S_{demerit}$: Magnitude of customer dissatisfaction, as perceived by the customer following the experience (observation)

$S_{merit}^{P}$: Magnitude of customer satisfaction, as perceived by the employee following the experience (observation)

$S_{demerit}^{P}$: Magnitude of customer dissatisfaction, as perceived by the employee following the experience (observation)

$x$: Magnitude of functional performance, reflecting the varying degrees of perceived customer satisfaction and dissatisfaction

$TS$: Gross social surplus

$PS$: Producer surplus

$CS$: Consumer surplus

$p^P$: Provider price

$p^R$: Recipient price

Business intelligence derived from service science and service engineering provides an effective way to empirically evaluate the variables defined above. Also, for readability, these definitions are restated as appropriate below.

\subsection{Overview of Fukuda (2022)}

This section presents an overview of Fukuda (2022), on which the present study is based. In essence, by applying the definition of service in the Arai–Shimomura model in the field of service engineering—an act whereby the service provider causes the change in state desired by the recipient in exchange for compensation (Shimomura et al., 2005)—to the definition employed in observation theory and performing computation based on quantum information theory, a mathematical structure is obtained that satisfies the properties of behavioral economics.

\subsubsection{Formulation of the Repeat Rate}

The churn rate and repeat rate are formulated by applying the definition of service in the Arai–Shimomura model in the field of service engineering to the definition employed in observation theory, as follows.

\begin{equation}
p(\uparrow) = \text{Tr} {\rho P(\uparrow)}, \quad p(\downarrow) = \text{Tr} {\rho P(\downarrow)},
\end{equation}

\begin{equation}
p(\uparrow) = \frac{1}{N} \sum_{n=0}^{N-1} |\alpha_n|^2 - \frac{2}{N^2 - N} \sum_{m>n} |\alpha_n^* \alpha_m|\cos\left((\theta_m - \phi_m) - (\theta_n - \phi_n)\right),
\end{equation}

\begin{equation}
p(\downarrow) = 1 - p(\uparrow),
\end{equation}

\begin{equation}
\sum_{n=0}^{N-1} |\alpha_n|^2 = t, \quad (0 \leq t \leq N).
\end{equation}

In the above, $p(\uparrow)$ = 0.15 and $p(\downarrow)$ = 0.85  serve as the neutral baseline for a biased state. To define a baseline free of any particular bias, the first moment (expected value, 0.5) and the square root of the second moment (root-mean-square value, $\frac{1}{\sqrt{2}}$)—two of the most fundamental and universal measures in statistics—are adopted as representative values.

\subsubsection{Rationale for Adopting the Neutral Baseline}

•	Assume that satisfaction/dissatisfaction follows a Bernoulli distribution in classical probability theory. Then, as in an unbiased coin toss, the expected value converges as trials are sufficiently repeated (by the law of large numbers):

\begin{equation}
|\alpha_n|^2=\frac{1}{2}.
\end{equation}

•	When $N=2$, a single experience of an intangible good is simply a rotation on the Bloch sphere satisfying $\cos\left((\theta_m - \phi_m) - (\theta_n - \phi_n)\right)=\frac{1}{\sqrt{2}}$. Also, consider the interference strength arising from an unbiased random phase, which follows a uniform distribution. The relevant quantity is the standard deviation—the standard measure of dispersion in statistics—and, by analogy, the root-mean-square value obtained from the mean wave energy in physics or electrical circuits:

\begin{equation}
\cos\left((\theta_m - \phi_m) - (\theta_n - \phi_n)\right)=\frac{1}{\sqrt{2}}.
\end{equation}

•	It is also numerically consistent with the $2$:$8$ rule (Pareto’s principle).

•	With $n$ denoting the number of repeats, the repeat rate is given by

\begin{equation}
p_{n+1}{(\downarrow)}=0.8-0.6p_{n}{(\downarrow)}+q_{n+1}.
\end{equation}

\subsubsection{Formulation of Customer Satisfaction}

The value functions for customer satisfaction and customer dissatisfaction, derived from the churn rate and the repeat rate using Shannon entropy, are expressed as follows:

\begin{equation}
S_{merit} = p(\uparrow)\log\left(\frac{x}{p(\uparrow)}+1\right), \quad 
S_{demerit} = p(\downarrow)\log\left(\frac{x}{p(\downarrow)}+1\right).
\end{equation}

\subsubsection{Properties of Customer Satisfaction}

•	Entropy and the utility function are equivalent (proven based on the minimum condition for the Kullback–Leibler divergence).

•	The value function encompasses the one in prospect theory.

•	The independence axiom is violated (though not in the case of the value function in prospect theory).

•	The functional form coincides with the specification used in econometrics.

•	The value functions for the case of $p(\uparrow)=0.2$ and $p(\downarrow)=0.8$, which are close to the neutral baseline for a biased state, are expressed as 

\begin{equation}
S_{merit} = 0.2\log\left(\frac{x}{0.2}+1\right), \quad x>0; \quad 
S_{demerit} =-0.8\log\left(-\frac{x}{0.8}+1\right), \quad x<0.
\end{equation}

Notably, their functional form mostly coincides with that of the value function in prospect theory.

\subsection{Overview of This Study}

An overview of this paper is given below.

From $p(\uparrow) = \text{Tr}{\rho P(\uparrow)}$ in Fukuda (2022), the fourfold composite system, comprising the provider’s and recipient’s perceptions about both intangible goods and the environment (rules), is formulated as a tensor product based on quantum information theory as follows:

\begin{equation}
p(\uparrow) = \text{Tr} \left( \rho_{P} \otimes \rho_{PE} \otimes \rho_{R} \otimes \rho_{RE} \right) \left( P_{P}(\uparrow) \otimes P_{PE}(\uparrow) \otimes P_{R}(\uparrow) \otimes P_{RE}(\uparrow) \right),
\end{equation}

Taking the partial trace corresponding to the recipient’s perception about the intangible good yields the following:

\begin{equation}
p_{r}(\uparrow) = \text{Tr} \, \text{Tr}_{RE}\left( \rho_{R} \otimes \rho_{RE} \right)\left(P_{R}(\uparrow) \otimes P_{RE}(\uparrow) \right).
\end{equation}

The POVM represents the perception about the intangible good after the experience. Requiring entanglement in the POVM and expanding the expression yields the following:

\begin{equation}
\begin{aligned}
p_{r}\left(\uparrow\right) = \frac{1}{N^2} \sum_{n=0}^{N-1} |\alpha_{rn}\alpha_{ren}|^2 - \frac{2}{N^2-N} \sum_{m>n} |\alpha_{rn}\alpha_{ren}\alpha_{rm}^*\alpha_{rem}^*| \cos\left( (\theta_{rm} + \theta_{rem} - \phi_{rm} - \phi_{rem}\right) \\ - \left(\theta_{rn} + \theta_{ren} - \phi_{rn} - \phi_{ren}) \right),
\end{aligned}
\end{equation}

\begin{equation}
p(\downarrow) = 1 - p(\uparrow),
\end{equation}

\begin{equation}
\sum_{n=0}^{N-1} |\alpha_{rn}\alpha_{ren}|^2= t, \quad (0 \leq t \leq N^2).
\end{equation}

\subsection{Theoretical Background of Fukuda (2022)}

This section describes Fukuda (2022), on which this study is based.

\subsubsection{Representation of Intangible Goods in the Hilbert Space}

Among studies analyzing intangible goods by decomposing them into processes, Nagashima and Nishio (2013) model retail services provided by online stores and consumer electronics retailers by dividing the customer’s experience, from entry to exit, into processes. They then compile questionnaire-based scores, on a 100-point scale, for each process and for the outcome, and thereby measure the impact of these factors. The results of the customer questionnaire on retail services verify that the functional form used in prospect theory is suited for a model explaining the overall evaluation in terms of the process evaluation and outcome evaluation. With this result in mind, Fukuda (2022) sets up a mathematical model.

A Hilbert space with processes as its basis is considered. This makes it theoretically possible to treat intangible goods scientifically by exploiting the way the basis in the Hilbert space is chosen. When multiple intangible goods of the same kind can be divided into the same kinds of processes, these intangible goods can be represented in a single Hilbert space. Since the basis coefficients differ across customers, the direction of the vector likewise varies from one customer to another. This represents the subjectivity of customers’ perceptions about intangible goods.

\subsubsection{Model Based on Observation Theory}

A system in which a customer experiences an intangible good (hereinafter, an experience system) is conceptualized from the standpoint of observation theory: the customer is regarded as the observer, the intangible good as the object of observation, and the customer’s experience of the intangible good as the act of observation. An experience system for an intangible good consisting of two processes is then examined. For example, ordering and dining are the two processes constituting the business of restaurants such as McDonald’s. The two processes covering a customer’s activities from entry to exit at McDonald’s are shown in Figure 1.

\begin{figure}[h]
\centering
\includegraphics[width=10cm]{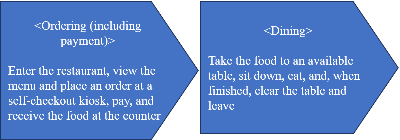}
\caption{Two processes at McDonald’s.}
\end{figure}

Each process (ordering and dining in the example) is assigned a weight representing the impact of the perception experienced. Also, it is theoretically possible to divide the two processes further to express them in greater detail.

Next, Dirac notation is introduced for the basis vectors representing the processes. For convenience, they are denoted by numbers, with ordering as $|0\rangle$  and dining as $|1\rangle$. The notation expressed in two-dimensional matrix form is as follows:

\begin{equation}
\langle 0 | = \begin{pmatrix} 1 & 0 \end{pmatrix}, \quad 
\langle 1 | = \begin{pmatrix} 0 & 1 \end{pmatrix}, \quad 
| 0 \rangle = \begin{pmatrix} 1 \\ 0 \end{pmatrix}, \quad 
| 1 \rangle = \begin{pmatrix} 0 \\ 1 \end{pmatrix}.
\end{equation}

In general, $\langle n-1 |$ is the unit row vector with 1 in the $n$th column, and $| n-1 \rangle$  is the unit column vector with 1 in the $n$th row.

The intangible good, represented by the two separate processes experienced (observed) by the customer, is denoted by $|\pi_0\rangle$ and $|\pi_1 \rangle$, which are expressed as 

\begin{equation}
|\pi_0\rangle=b_0| 0 \rangle+b_1| 1 \rangle, \quad |\pi_1\rangle=b_{1}^{*}| 0 \rangle-b_{0}^{*} |1\rangle.
\end{equation}

$|b_{0}|^2$ and $|b_{1}|^2$ represent the weights for the customer’s experience (observation) of the processes $|0\rangle$ and $|1\rangle$, respectively, and are assumed to satisfy the normalization condition $|b_{0}|^2+|b_{1}|^2=1$. Also, under $\langle \pi_1|=b_1\langle0|-b_0\langle1|$, $\langle \pi_1|\pi_0\rangle=0$ holds. It is then assumed that observation brings $|\pi_0\rangle$  into the states $|+\rangle$ and $|-\rangle$ expressed below, and that this reproduces the actual single experience process seen in Figure 2.

\begin{equation}
|+\rangle=\alpha| 0 \rangle-\beta| 1 \rangle, \quad |-\rangle=\beta^{*}| 0 \rangle+\alpha^{*} |1\rangle.
\end{equation}

$b_0$, $b_1$, $\alpha$, and $\beta$ are complex numbers; $b_0^{*}$, $b_1^{*}$, $\alpha^{*}$, and $\beta^{*}$ are their respective complex conjugates. In polar form, they are expressed as $b_0=|b_0|$, $b_1=|b_1|e^{i\theta}$, $\alpha=|\alpha|$, and $\beta=|\beta|e^{i\phi}$; $i$  is the imaginary unit, $i=\sqrt{-1}$; and $\theta$  and $\phi$  are real numbers.

By letting $p(\uparrow)$ and $p(\downarrow)$ denote the respective weights for the state of the intangible good changing to $|+\rangle$  and $|-\rangle$ when the customer experiences (observes) the intangible good $|\pi_0\rangle$, the following is obtained:

\begin{equation}
p(\uparrow)=|\langle\pi_0|+\rangle|^2 = |b_0|^{2}|\alpha|^{2}+|b_1|^{2}|\beta|^{2}-2|b_{0}b_{1}\alpha\beta|\cos(\theta-\phi).
\end{equation}

The equation below is obtained in a similar manner:

\begin{equation}
p(\downarrow)=|\langle\pi_0|-\rangle|^2 = |b_0|^{2}|\alpha|^{2}+|b_1|^{2}|\beta|^{2}+2|b_{0}b_{1}\alpha\beta|\cos(\theta-\phi).
\end{equation}

From $p(\uparrow)+p(\downarrow)=1$, $|b_0|^{2}|\alpha|^{2}+|b_1|^2 |\beta|^2=\frac{1}{2}$ is obtained, which yields

\begin{equation}
p(\uparrow)=\frac{1}{2}-2|b_{0}b_{1}\alpha\beta|\cos(\theta-\phi),
\end{equation}

\begin{equation}
p(\downarrow)=\frac{1}{2}+2|b_{0}b_{1}\alpha\beta|\cos(\theta-\phi). \end{equation}

The second term on the right-hand side corresponds to the interference term in quantum mechanics.

\subsubsection{Interpretation of the Weights Following the Experience (Observation) of the Intangible Good}

By assigning each process experienced (observed) by the customer its average weight, the following is obtained:

\begin{equation}
|b_0|^2=|b_1|^2=\frac{1}{2}.
\end{equation}

The POVM parameters are defined as 

\begin{equation}
|\alpha|^2=|\beta|^2=\frac{1}{2}.
\end{equation}

Substituting these yields

\begin{equation}
p(\uparrow)=\frac{1}{2}-\frac{1}{2}\cos(\theta-\phi),
\end{equation}

\begin{equation}
p(\downarrow)=\frac{1}{2}+\frac{1}{2}\cos(\theta-\phi).
\end{equation}

Taking the root-mean-square value as the average gives $\cos(\theta-\phi)=\frac{1}{\sqrt{2}}$, and substituting this yields the following estimates:

\begin{equation}
p(\uparrow)=0.15, \quad p(\downarrow)=0.85 .
\end{equation}

\begin{figure}[h]
\centering
\includegraphics[width=10cm]{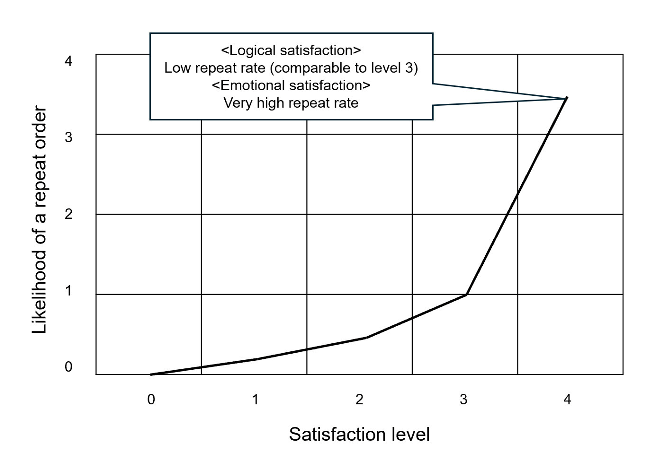}
\caption{Emotional satisfaction as a driver of repeat orders. Partially modified from Suwa (2015).}
\end{figure}

As Figure 2 illustrates satisfaction levels progressing from low to high along the x-axis, the weight for dissatisfaction is defined by the width ratio from the left, while the weight for satisfaction corresponds to the remaining width ratio. In other words, the weight ratio of satisfaction to dissatisfaction is $0.2$:$0.8$, which closely approximates the theoretical value.
The interpretation of $p(\uparrow)$ and $p(\downarrow)$ is provided as follows.

Regarding satisfaction, let us first consider logical satisfaction. The logical satisfaction term represents the outcome of experiencing the intangible good, where a fair coin toss analogy is objectively applied to the customer’s perceived satisfaction or dissatisfaction, and the weight for either satisfaction or dissatisfaction becomes $\frac{1}{2}$. This term also represents the expected value in classical probability theory for the occurrence of either satisfaction or dissatisfaction. Interpreted as a fair coin-toss, the term converges to $\frac{1}{2}$ over a sufficiently large number of trials according to the law of large numbers. Hence, the term is referred to as the logical satisfaction term. Next, let us consider emotional satisfaction. $\frac{\cos(\theta-\phi)}{2}$ is treated as a correction to the logical satisfaction term; as it pertains to individual processes, it is referred to as the emotional satisfaction term. Based on the above, $p(\uparrow)$ and $p(\downarrow)$ denote the weights of emotional satisfaction that affect the repeat rate; specifically, $p(\uparrow)$ and $p(\downarrow)$ are defined as the weights for the satisfaction and dissatisfaction factors, respectively. Processes are components of an intangible good, and its functions manifest through the use of, or engagement in, these processes. Here, it is assumed that a single function manifests in response to the use of, or engagement in, a single process. That is, a process corresponds to a function. Consequently, $p(\uparrow)$ and $p(\downarrow)$—defined as the weight ratios for the satisfaction and dissatisfaction factors, respectively—can also be viewed as representing the weights for functional performance levels (high, low), as well as the usage levels of the intangible good (effectively utilized, underutilized).

This can also be explained by introducing $|\pi_1\rangle$ satisfying $\langle\pi_1|\pi_0\rangle=0$ and assuming that observation brings $|\pi_0\rangle$ into the states $|+\rangle$ and $|-\rangle$, which are obtained by rotation on the Bloch sphere as shown below. This is taken to reproduce the actual single experience process seen in Figure 2. In other words, with the global phase given by $e^{i\lambda}=-i$, a single experience of an intangible good is simply a rotation on the Bloch sphere satisfying $\cos{2\psi}=\frac{1}{\sqrt{2}}$, as shown below.

\begin{equation}
|+\rangle=\sin\psi|\pi_0\rangle+i\cos\psi|\pi_1\rangle,
\end{equation}

\begin{equation}
|-\rangle=\cos\psi|\pi_0\rangle-i\sin\psi|\pi_1\rangle,
\end{equation}

\begin{equation}
|\pi_0\rangle=\frac{1}{\sqrt{2}}|0\rangle+\frac{1}{\sqrt{2}}|1\rangle,
\end{equation}

\begin{equation}
|\pi_1\rangle=\frac{1}{\sqrt{2}}|0\rangle-\frac{1}{\sqrt{2}}|1\rangle,
\end{equation}

\begin{equation}
\psi=\frac{\theta-\phi}{2},
\end{equation}

\begin{equation}
p(\uparrow)=|\langle\pi_0|+\rangle|^2=\sin^{2}\psi=\frac{1}{2}-\frac{1}{2}\cos2\psi,
\end{equation}

\begin{equation}
p(\downarrow)=|\langle\pi_0|-\rangle|^2=\cos^{2}\psi=\frac{1}{2}+\frac{1}{2}\cos2\psi.
\end{equation}

For clarity, the redundant phase preceding $|1\rangle$ is set to zero so that $|\pi_0\rangle$ and $|\pi_1\rangle$ form the x-basis on the Bloch sphere. Furthermore, the perception about the intangible good, $|\pi_0\rangle$, can be transformed into the following.

\begin{equation}
|\pi_0\rangle=\sin\psi|+\rangle+\cos\psi|-\rangle.
\end{equation}

Therefore, it can be expressed as a quantum state, a superposition of the customer’s perceptions of satisfaction and dissatisfaction after experiencing the intangible good. Although this quantum state is derived for an intangible good consisting of two processes, it can also be shown to represent one consisting of an arbitrary number of processes.

\subsubsection{Value Function for Customer Satisfaction}

In service engineering research, a service is defined as an act whereby the service provider causes the change in state desired by the recipient in exchange for compensation (Shimomura et al., 2005). The present paper regards such a service as an intangible good. As before, an experience system for an intangible good is conceptualized from the standpoint of observation theory: the customer is regarded as the observer, the intangible good as the object of observation, and the customer’s experience of the intangible good as the act of observation. Here, an experience system for an intangible good consisting of an arbitrary number of processes is examined. 

The intangible good, which is expressed by a sequence of $N$ processes experienced (observed) by the customer is represented by $|\pi\rangle$ as follows.

\begin{equation}
|\pi\rangle=\sum_{n=0}^{N-1}b_n|n\rangle,
\end{equation}

where $|\pi\rangle$  assumed to satisfy the normalization condition. The density matrix $\rho$  representing the state of the intangible good is given as follows.

\begin{equation}
\rho=|\pi\rangle\langle\pi|.
\end{equation}

Next, for $|+\rangle=\sum_{n=0}^{N-1}\alpha_{n}|n\rangle$ and $|-\rangle=\sum_{n=0}^{N-1}\beta_{n}|n\rangle$, a POVM, which is a mathematical representation of a measurement, is defined as follows. Here, it is not required that $|+\rangle$ and $|-\rangle$ satisfy the normalization or orthogonality conditions.

\begin{equation}
P(\uparrow)=|+\rangle\langle+|, \quad P(\downarrow)=|-\rangle\langle-|. 
\end{equation}

Based on the assumptions of the POVM, the following is obtained.

\begin{equation}
P(\uparrow)+P(\downarrow)=I,\notag
\end{equation}

\begin{equation}
\sum_{n=0}^{N-1}(|\alpha_n|^2+|\beta_n|^2)|n\rangle\langle n|+\sum_{m \ne n}(\alpha_{m}^{*}\alpha_{n}+\beta_{m}^{*}\beta_n)|m\rangle\langle n|=I,\notag
\end{equation}

\begin{equation}
\alpha_{m}^{*}\alpha_{n}+\beta_{m}^{*}\beta_n=\delta_{mn}. 
\end{equation}

Here, $I$ denotes the identity matrix, $\delta_{mn}$ represents the Kronecker delta (which equals $1$ if $m=n$ , and $0$ if $m \ne n$ ), and $\sum_{m \ne n}$ denotes the summation over all terms satisfying the condition $m \ne n$.

If the weights for the state of the intangible good transitioning into $P(\uparrow)$ or $P(\downarrow)$ upon the customer’s experience (observation) are denoted by $P(\uparrow)$ and $P(\downarrow)$, respectively, the following is obtained.

\begin{equation}
p(\uparrow)=\text{Tr}\rho P(\uparrow), \quad p(\downarrow)=\text{Tr}\rho P(\downarrow).
\end{equation}

Next, letting $b_n=|b_n|e^{i\theta_n}$ and $\alpha_n=|\alpha_n|e^{i\phi_n}$, and setting $|b_n|^2=\frac{1}{N}$ for simplicity, the following is obtained.

\begin{equation}
p(\uparrow)=\frac{1}{N}\sum_{n=0}^{N-1}|\alpha_n|^2-\frac{2}{N}\sum_{m>n}|\alpha_n^{*}\alpha_m|\cos((\theta_m-\phi_m)-(\theta_n-\phi_n)),
\end{equation}

\begin{equation}
p(\downarrow)=1-p(\uparrow), 
\end{equation}

where $\sum_{m>n}$ denotes the summation over all terms satisfying the condition $m>n$ .

Next, equations (40) and (41) are reinterpreted, relaxing the condition regarding $m=n$  in equation (38) and the constraint on the first term of the right-hand side of equation (40). Now, by letting $\sum_{n=0}^{N-1}|\alpha_n|^2=t\,(0\leq t\leq N)$, the satisfaction level evaluated for each process/function can be represented as a portion of $t$. Normalizing the second term on the right-hand side by the number of combinations of two from $N$, ${}_N C_2=\frac{N^2-N}{2}$, yields the following.

\begin{equation}
p(\uparrow)=\frac{1}{N}\sum_{n=0}^{N-1}|\alpha_{n}|^{2}-\frac{2}{N^{2}-N}\sum_{m>n}|\alpha_{n}^{*}\alpha_{m}|\cos((\theta_{m}-\phi_{m})-(\theta_{n}-\phi_{n})),
\end{equation}

\begin{equation}
p(\downarrow)=1-p(\uparrow). 
\end{equation}

$\sum_{m>n}$ denotes the summation over all terms satisfying the condition $m>n$. As before, in terms of psychological meaning, the first term on the right-hand side of equation (42) is the logical satisfaction term, corresponding to the expected value in classical probability theory; the second term on the right-hand side is the emotional satisfaction term, which serves as a correction term representing emotion; and the phase represents service quality. The entropy of the experience system for an intangible good diffused to level $x$ is then obtained from the per-unit experience system as follows:

\begin{equation}
S = \left( - \frac{p\left(\uparrow \right)}{x}\log \left(\frac{p \left(\uparrow \right)}{x} \right) - \frac{p \left(\downarrow \right)}{x}\log \left(\frac{p \left(\downarrow \right)}{x}\right) \right)x \notag
\end{equation}

\begin{equation}
= p\left(\uparrow\right)\log \left(\frac{x}{p\left(\downarrow\right)}\right) + p\left(\downarrow\right)\log \left(\frac{x}{p\left(\downarrow\right)}\right) \notag
\end{equation}

\begin{equation}
=S_{merit}+S_{demerit}.
\end{equation}

Shifting the curves to ensure consistency at $x=0$  yields the following nonlinear asymmetric value functions representing gains and losses:

\begin{equation}
S_{merit} = p\left(\uparrow\right)\log \left(\frac{x}{p\left(\uparrow\right)}+1\right), \quad
S_{demerit} = p\left(\downarrow\right)\log \left(\frac{x}{p\left(\downarrow\right)}+1\right).
\end{equation}

To discuss bias, this paper considers a neutral baseline regarding the way in which bias occurs. Satisfaction/dissatisfaction in the logical satisfaction term of $p(\uparrow)$ and $p(\downarrow)$ is assumed to follow a Bernoulli distribution in classical probability theory, as in an unbiased coin toss. Then, with a sufficient number of repeated trials (or experiences of an intangible good), the expected value converges to $|\alpha_n|^2=\frac{1}{2}$ by the law of large numbers. Also, when $N=2$, a single experience of an intangible good is known to be simply a rotation on the Bloch sphere satisfying $\cos{2\psi}=\frac{1}{\sqrt{2}}$, with the global phase given by $e^{i\lambda}=-i$. Additionally, the phase in the emotional satisfaction term of $p(\uparrow)$ and $p(\downarrow)$ is assumed to be random and unbiased, following a uniform distribution. Then, the standard deviation—the standard measure of dispersion in statistics—is given by $\cos((\theta_m-\phi_m)-(\theta_n-\phi_n))=\frac{1}{\sqrt{2}}$. By analogy, this is also consistent with the root-mean-square value obtained from the mean wave energy in physics or electrical circuits. The theoretical value of the weight ratio of $p(\uparrow)$ to $p(\downarrow)$ is then estimated to be $0.15$:$0.85$. As shown below, transforming this entropy gives a form identical to the value function in prospect theory (Figure 3):

\begin{equation}
S_{merit}=0.15\log\left(\frac{x}{0.15}+1\right), \quad x>0, 
\end{equation}

\begin{equation}
S_{demerit}=-0.85\log\left(-\frac{x}{0.85}+1\right), \quad x<0.
\end{equation}

In particular, as shown in the Appendix, when this weight ratio is $0.2$:$0.8$, the entropy has been proven to be equivalent to the value function in prospect theory, confirming that the neutral baseline is appropriate. Also, this entropy, which retains the asymmetry, is maximized when the logical satisfaction term is held fixed and two of the processes carry large weights in the emotional satisfaction term. In other words, the peak-end rule holds. In addition, the weight ratio of $0.2$:$0.8$ reflects Pareto’s principle.

\begin{figure}[h]
\centering
\includegraphics[width=10cm]{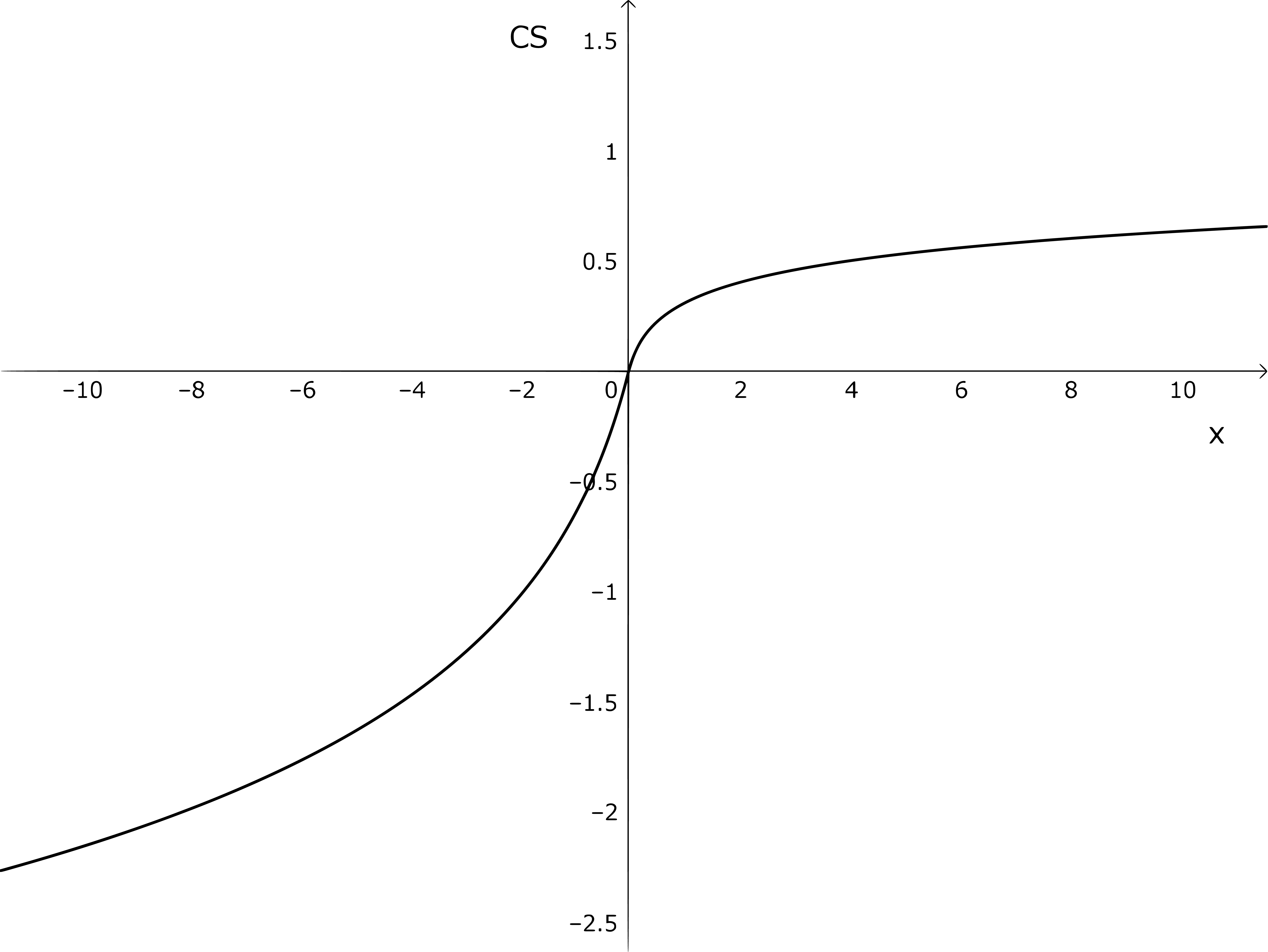}
\caption{Graph of customer satisfaction. Here, $x$ is a measure of the prevalence of intangible goods in the economy. The right line represents $S_{merit} = 0.15 \log \left(\frac{x}{0.15} + 1\right)$, and the left line represents  $S_{demerit} = -0.85 \log \left(-\frac{x}{0.85} + 1\right)$.}
\end{figure}

While standard “goods” are analyzed using classical probability theory, the intangible goods considered in this paper are examined through the lens of quantum information theory. Consequently, the mathematical distinction between the classical conception of goods and the intangible goods presented here arises from the existence of an emotional satisfaction term (process), which corresponds to the interference term in quantum mechanics. The presence of this term causes a departure from the rational properties associated with standard goods and gives rise to irrational characteristics. In other words, the irrational properties identified in behavioral economics—including the asymmetry of the value function in prospect theory and the peak-end rule—stem from the second term on the right-hand side of the second equation below.

\begin{equation}
\quad p(\uparrow)=\frac{1}{N}\sum_{n=0}^{N-1}|\alpha_{n}|^2-\frac{2}{N^2-N}\sum_{m>n}|\alpha_{n}^{*}\alpha_{m}|\cos((\theta_{m}-\phi_{m})-(\theta_{n}-\phi_{n})).
\end{equation}

Since standard goods are modeled using classical probability theory, the quantum mechanical interference term is absent, as seen in the following equation.

\begin{equation}
p(\uparrow)=\frac{1}{N}\sum_{n=0}^{N-1}|\alpha_n|^2.
\end{equation}

For decision utility in standard economics, $S_{merit}^{dec}$,

\begin{equation}
p(\uparrow)=\frac{1}{N}\sum_{n=0}^{N-1}|\alpha_n|^2.
\end{equation}

For experienced utility in behavioral economics, $S_{merit}^{exp}$, 

\begin{equation}
p(\uparrow)=\frac{1}{N}\sum_{n=0}^{N-1}|\alpha_{n}|^2-\frac{2}{N^2-N}\sum_{m>n}|\alpha_{n}^{*}\alpha_{m}|\cos((\theta_{m}-\phi_{m})-(\theta_{n}-\phi_{n})),
\end{equation}

\begin{equation}
p(\downarrow)=1-p(\uparrow),
\end{equation}

\begin{equation}
\sum_{n=0}^{N-1}|\alpha_{n}|^2=t, \quad\left(0\leq t \leq N\right).
\end{equation}

Internality $\Lambda$ is then expressed as follows:

\begin{equation}
\Lambda=S_{merit}^{dec}-S_{merit}^{exp}.
\end{equation}

\subsubsection{Formulation of the Repeat Rate}

This section considers an expression for the repeat rate obtained by adding a correction term to the Bayesian expression (Takahashi, 2013; Yukalov and Sornette, 2017). The weight for emotional satisfaction that affects the repeat rate is expressed as

\begin{equation}
p_{n+1}(\uparrow)=p_{n+1|n}(\uparrow|\uparrow)p_{n}(\uparrow)+p_{n+1|n}(\uparrow|\downarrow)p_{n}(\downarrow)-q_{n+1}. 
\end{equation}

Here, $n$ denotes the number of repeats, and $p_{n}(\uparrow)$ and $p_{n}(\downarrow)$ represent, respectively, the weights for satisfaction and dissatisfaction at the $n$th repeat of the intangible good. $p_{n+1|n}(\uparrow|\uparrow)$ and $p_{n+1|n}(\uparrow|\downarrow)$ represent the weights for satisfaction at the $(n+1)$ th repeat, given that the $n$th repeat resulted in satisfaction and in dissatisfaction, respectively. $q_{n+1}$ denotes the correction term representing emotional satisfaction at the $(n+1)$ th repeat. From Figure 2, setting $p_{n+1|n}(\uparrow|\uparrow)=0.2$  and $p_{n+1|n}(\uparrow|\downarrow)=0.8$  gives the following:

\begin{equation}
p_{n+1}(\uparrow)=0.2p_{n}(\uparrow)+0.8p_{n}(\downarrow)-q_{n+1}. 
\end{equation}

When $p_{n}(\uparrow)=0.2$, $p_{n}(\downarrow)=0.8$ holds, and reducing $p_{n}(\uparrow)$ to raise the level of satisfaction increases $p_{n}(\downarrow)$. Figure 2 also shows that $p_{n}(\downarrow)$ corresponds to the level of satisfaction. Accordingly, the repeat rate is expressed as follows, taking as its indicator the weight for dissatisfaction $p_{n}(\downarrow)$, which represents loss in Figure 2.

\begin{equation}
p_{n+1}(\downarrow)=0.8-0.6p_{n}(\downarrow)+q_{n+1}. 
\end{equation}

For $n=0$ , representing a new customer, setting $p_{0}(\downarrow)=0.8$ and $q_1=0.35$ as average values gives an average repeat rate $\overline{p_{1}(\downarrow)}$ of $0.67$. Also, the repeat rates at satisfaction levels 3 and 4 in Figure 2, to which this equation applies, are $0.25$ and $0.9$, respectively, and their average of 0.58 lies close to the theoretical value. Furthermore, setting $p_{0}(\downarrow)=0.6$ and $q_{1}=0.35$  for satisfaction levels 3 and 4 in Figure 2 gives an average repeat rate $\overline{p_{1}(\downarrow)}$ of $0.79$, which indicates satisfaction level 4, where most customers repeat.

Lastly, the value function for customer satisfaction $S_{merit}$, expressed in terms of the repeat rate $p(\downarrow)$, is as follows.

\begin{equation}
S_{merit}=(1-p(\downarrow))\log\left(\frac{x}{1-p(\downarrow)}+1\right).
\end{equation}

\subsection{Perception of Intangible Goods Considering the Environment}

According to Fukuda (2023, 2024), tensor products are used to represent a composite system of multiple individuals in quantum information theory. From $p(\uparrow)=\text{Tr}\rho P(\uparrow)$, the fourfold composite system, comprising the provider’s and recipient’s perceptions of both intangible goods and the environment (rules), is formulated as a tensor product based on quantum information theory as follows.

\begin{equation}
p(\uparrow) = \text{Tr} \left( \rho_{P} \otimes \rho_{PE} \otimes \rho_{R} \otimes \rho_{RE} \right) \left( P_{P}(\uparrow) \otimes P_{PE}(\uparrow) \otimes P_{R}(\uparrow) \otimes P_{RE}(\uparrow) \right),
\end{equation}

where \(\rho_{P}\), \(\rho_{PE}\), \(\rho_{R}\), and \(\rho_{RE}\)  are density matrices; \(
P_{P}(\uparrow)\), \(P_{PE}(\uparrow)\), \(
P_{R}(\uparrow)\), and \(P_{RE}(\uparrow)\) are POVMs; \(\rho_{P}\) and \(
\rho_{R}\) respectively represent the provider's and the recipient's perception of the provision/receipt process before the actual provision/receipt of intangible goods; \(\rho_{PE}\) and \(\rho_{RE}\) respectively represent the provider's and the recipient's perception of the environment (rules) before the actual provision/receipt of intangible goods; \(
P_{P}(\uparrow)\)  and \(P_{R}(\uparrow)\) respectively represent the provider's and the recipient's perception of the process after the actual provision/receipt of intangible goods; and \(P_{PE}(\uparrow)\) and  \(P_{RE}(\uparrow)\)  respectively represent the provider's and the recipient's perception of the environment (rules) after the actual provision/receipt of intangible goods.

The discussion regarding equation (59) follows the same logical progression.

In line with the model by Fukuda (2022), the above general expressions are assumed for the product of weights $(p(\downarrow))$ for emotional satisfaction with intangible goods and rules (i.e., the environment) that affects the provider’s and the recipient’s repeat rates, when the recipient’s repeat rate changes after the provision or receipt of intangible goods.

Considering the case where the provider's and the recipient's perceptions of intangible goods are correlated with their respective perceptions of the environment (rules) after the actual provision/receipt of intangible goods, the product of weights  $(p(\uparrow))$ for emotional satisfaction with intangible goods and the environment (rules), which affects the provider's and recipient's repeat rates, is expressed as follows:

\begin{equation}
p(\uparrow) = \text{Tr}\left(\rho_{P}\otimes \rho_{PE}\right)\left(P_{P}(\uparrow)\otimes P_{PE}(\uparrow)\right)\text{Tr}\left(\rho_{R}\otimes \rho_{RE}\right) \left(P_{R}(\uparrow) \otimes P_{RE}(\uparrow)\right).
\end{equation}

In this case, with regard to the provision and receipt processes, which are components of the perception of intangible goods and the environment (rules) after the actual provision/receipt of intangible goods, the same processes are related to each other, whereas different processes are not. Therefore, the POVMs that require entanglement between the state of perception of intangible goods and the state of perception of the environment (rules), as mutually corresponding interactions, are set up as follows.

\begin{equation}
\rho_{R}\otimes\rho_{RE} = \sum_{i,j,i',j'}b_{ri}b_{rei'}b_{rj}^*b_{rej'}^*|ii'\rangle\langle jj'|,
\end{equation}

\begin{equation}
P_{R}(\uparrow)\otimes P_{RE}(\uparrow) \rightarrow \, \sum_{i,j}\alpha_{ri}\alpha_{rei}\alpha_{rj}^*\alpha_{rej}^*|ii\rangle\langle jj|,
\end{equation}

where $|ij\rangle = |i\rangle \otimes |j\rangle$, and the $(2i+j)$-th row is a unit vector. 

Then, the weight $\left(
p_{r}(\uparrow)\right)$ for emotional satisfaction that affects the provider's repeat rate is expressed as follows:

\begin{equation}
\begin{aligned}
p_{r}(\uparrow) = \text{Tr} \, \text{Tr}_{RE}\left( \rho_{R} \otimes \rho_{RE} \right)\left(P_{R}(\uparrow) \otimes P_{RE}(\uparrow) \right) \\ \rightarrow  \text{Tr} \, \text{Tr}_{RE}\sum_{i,j,i',j'}b_{ri}b_{rei'}b_{rj}^*b_{rej'}^*|ii'\rangle\langle jj'|
\sum_{i,j}\alpha_{ri}\alpha_{rei}\alpha_{rj}^*\alpha_{rej}^*|ii\rangle\langle jj| \\= \sum_{i,k}b_{ri}b_{rei}b_{rk}^*b_{rek}^*\alpha_{rk}\alpha_{rek}\alpha_{ri}^*\alpha_{rei}^*.
\end{aligned}
\end{equation}

Here, $\text{Tr}_{RE}$ is the partial trace over RE, which corresponds to the state of R (the focus of our interest), in the composite system described by the tensor product Hilbert space $R \otimes RE$, and the following equation holds $\rho=\sum_{i,j,i',j'}\rho_{i,j,i',j'}|ij\rangle\langle i'j'|$,

\begin{equation}
\text{Tr}_{RE}\rho = \sum_{i,i'}{(\sum_{j}\rho_{i,j,i',j}})|i\rangle\langle i^{'}|.
\end{equation}

Then, if $b_{ri}=|b_{ri}|e^{i\theta_{ri}}$, $b_{rei}=|b_{rei}|e^{i\theta_{rei}}$, $\alpha_{ri}=|\alpha_{ri}|e^{i\phi_{ri}}$, $\alpha_{rei}=|\alpha_{rei}|e^{i\phi_{rei}}$ and $|b_{ri}|^2=|b_{rei}|^2=\frac{1}{N}$ (as an average) and if the second term on the right-hand side is normalized ${}_N C_2=\frac{N^2-N}{2}$ (the number of combinations of two elements chosen from $N$ elements), the following equation is obtained:

\begin{equation}
\begin{aligned}
p_{r}(\uparrow) = \frac{1}{N^2} \sum_{n=0}^{N-1} |\alpha_{rn}\alpha_{ren}|^2 - \frac{2}{N^2 - N} \sum_{m>n} |\alpha_{rn}\alpha_{ren}\alpha_{rm}^*\alpha_{rem}^*| \cos \left( (\theta_{rm} + \theta_{rem} - \phi_{rm} - \phi_{rem} \right) \\ - \left(\theta_{rn} + \theta_{ren} - \phi_{rn} - \phi_{ren}) \right),
\end{aligned}
\end{equation}

\begin{equation}
\sum_{n=0}^{N-1} |\alpha_{rn}\alpha_{ren}|^2= t, \quad (0 \leq t \leq N^2).
\end{equation}

\section{Mathematical Model of Nudges in This Study}

A customer repeating an action because of loss aversion implies the customer averting losses resulting from lost operational rationality (mathematical rationality) in making judgements and decisions due to biases arising in the customer's preferences and probabilistic judgements. 

Within the standard economic framework, for optimal nudges, there is a trade-off between correcting biased individuals' internalities and psychological costs of nudges to all individuals (Jimenez-Gomez, 2018). For the model of this study (Figure 4), a rapid increase in bias leads to perceived loss due to a rapid decrease in customer satisfaction, then to an increase in the repeat rate through loss aversion, and finally to greater effects of nudges. Therefore, the repeat rate equals the rate of behavioral improvement due to nudges and nudges in standard economics are equivalent to nudges considered in this study. 

\begin{figure}[h]
\centering
\includegraphics[width=10cm]{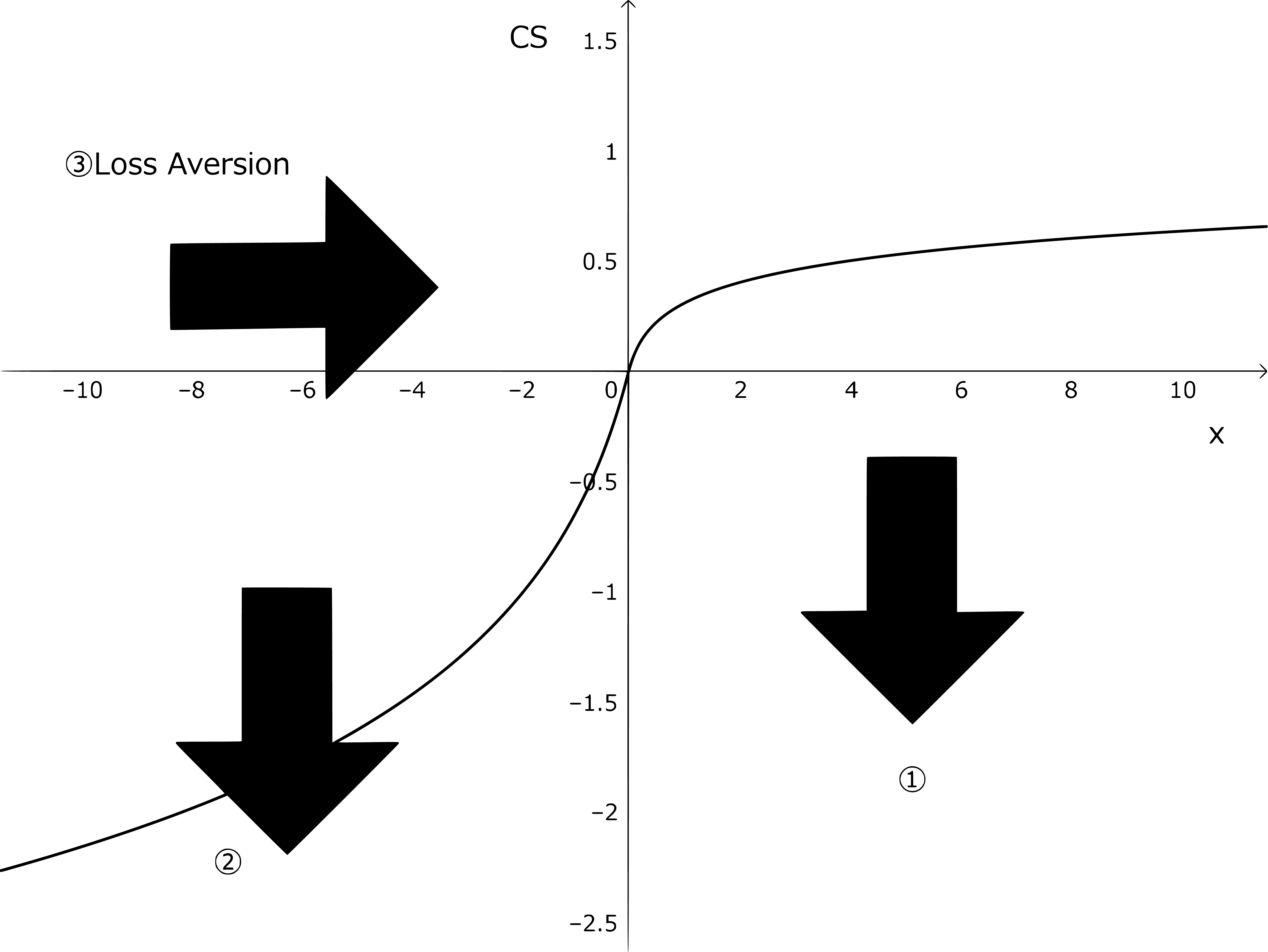}
\caption{Mathematical model of nudges. A rapid increase in bias (\textcircled{1}) leads to perceived loss due to a rapid decrease in customer satisfaction (\textcircled{1},\textcircled{2}), then to an increase in the repeat rate through loss aversion (\textcircled{3}), and finally to greater effects of nudges.}
\end{figure}

\section{Gross Social Surplus}

Let the recipient price $p^R$ and the provider price $p^P$ be positive and negative, respectively. Also, let the value function for customer satisfaction and the value function for employee satisfaction be $S_{merit}^R$ and $S_{merit}^P$ respectively. Let gross social surplus, producer surplus, and consumer surplus be $TS$, $CS$, and $PS$, respectively. Then, based on the first-order conditions for maximizing quasi-linear utility $(p^R = \frac{\partial S_{merit}^R}{\partial x}, \, p^P = -\frac{\partial S_{merit}^P}{\partial x})$, gross social surplus $TS$ is expressed as

\begin{equation} 
\begin{aligned}
TS = CS + PS = \int_0^x p^Rdx - \int_0^x p^Pdx = S_{merit}^R + S_{merit}^P 
\\=  p_{r}(\uparrow)\log\left(\frac{x}{p_{r}(\uparrow)}+1\right) + p_{p}(\uparrow)\log\left(\frac{x}{p_{p}(\uparrow)}+1\right).
\end{aligned}
\end{equation}

The emotional satisfaction term introduces bias and reduces welfare. A small emotional satisfaction term leads to high utility. A large emotional satisfaction term leads to a high repeat rate. Gross social surplus is greater than that in standard economics due to the presence of the emotional satisfaction term.

\section{Design of Customized Customer Experiences}

The customer's perception of intangible goods, $\rho$, changes to $P(\uparrow)$ by experiencing (observing) them, that is, \\$p(\uparrow)= \text{Tr}\rho P(\uparrow)$. Designing individually customized customer experiences based on customer insight is equivalent to determining the coefficients of the underlying factors in individual customers' subjective perception ($\rho$) of the process of receiving intangible goods that maximize the value function for customer satisfaction in accordance with the underlying factors and their coefficients for customers' perception ($P(\uparrow)$) of the process and, in some cases, in accordance with nudges (Figure 5).

\begin{figure}[h]
\centering
\includegraphics[width=10cm]{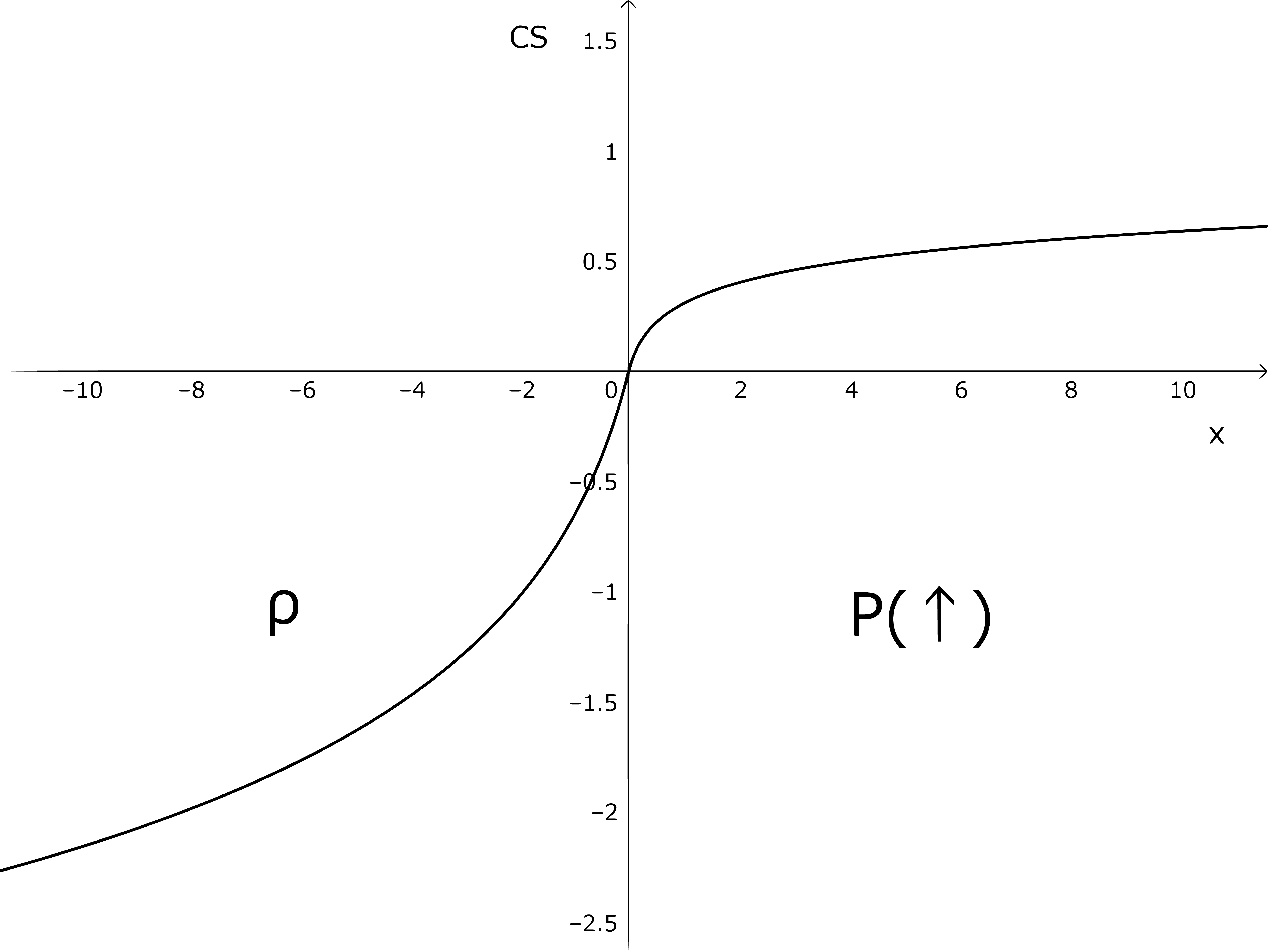}
\caption{Customer experiences. The customer's perception of intangible goods, $\rho$, changes to $P( \uparrow)$ by experiencing (observing) them.}
\end{figure}

\section{Methods for Validating Variables in This Study }

Business intelligence derived from service science and service engineering provides an effective way to empirically evaluate the variables defined above.

For instance, suppose that subjective data are collected by having customers self-report their perceived impact, prior expectations, and reputation on a 5-point scale via a questionnaire before and after they experience (observe) intangible goods through nudges. Alternatively, in cases where customers are present at a service site during a particular process, subjective survey data can be obtained by having them enter the level of impact they perceive—on a 5-point scale—into their smartwatch. Furthermore, as a common method of utilizing IT to collect objective data, multimodal learning—in which AI learns patterns in objective IT data using the pre- and post-experience (observation) subjective survey data as ground truth—enables the estimation of pre- and post-experience (observation) emotions. Specifically, emotions are estimated through multimodal learning, utilizing deep learning to recognize patterns within IT-based objective data. In this process, subjective survey data regarding impact, prior expectations, and reputation serve as the ground truth. The objective inputs include biological information and location and movement data collected through smartwatches, smartphones, or electroencephalographs, as well as data collected through communication systems (e.g., voice call, email, and text messaging systems) and on-site systems (e.g., LRF and LiDAR devices, RGB-D cameras). Measuring objective IT data from on-site customers in process $i$ allows for the empirical validation of a customer’s perceived impacts, expectations, and reputations regarding the intangible good and the environment (rules). These are treated as customized theoretical values captured both before and after the experience (observation). Next, the rate of behavioral change can be statistically measured based on the proportion of customers who repeat a purchase or exhibit a change in behavior following the experience (observation) through nudges.

Emotions can also be estimated through multimodal learning, utilizing the statistically measured behavioral change rate as ground truth and performing deep learning to recognize patterns within specific customers’ objective IT data. This makes it possible to validate the rate of behavioral change as a customized theoretical value, based on customers’ objective IT data. Then, adjusting $t$  in equations (53) and (66) allows for the validation of customized theoretical values for the behavioral change rate $(p_{r}(\downarrow))$ and churn rate $(p_{r}(\uparrow))$. These values are calculated using equations (50) through (52) based on measurements of impact, expectations, and reputation captured both prior to and following the experience. Furthermore, regarding the functional performance of the intangible good following the experience (observation), the customer is asked to identify a neutral point representing neither satisfaction nor dissatisfaction. The deviation from this reference point, denoted as $x$, indicates customer satisfaction when positive and dissatisfaction when negative. For example, in a questionnaire, customers are presented with seat pitch (the distance between their seat and the one in front) and asked to identify the specific length they consider neutral—neither satisfactory nor unsatisfactory. Lengths exceeding this reference point are assigned positive values, representing the satisfaction range, whereas lengths below it are assigned negative values, representing the dissatisfaction range. Then, subjective data can be measured by having customers respond to a five-point scale questionnaire that includes questions on their level of satisfaction or dissatisfaction and, for satisfied customers, questions on consumer surplus and the recipient price. In this case, a 5-point scale consisting of 0, 1, 2, 3, and 4 is mapped to the following ranges of the rate $p_{r}(\downarrow)$: $0$–$0.2$, $0.2$–$0.4$, $0.4$–$0.6$, $0.6$–$0.8$, and $0.8$–$1$, respectively. The level of customer satisfaction or dissatisfaction can then be determined as a function of $x$ according to equation (45). Given $x$, emotions are estimated through multimodal learning; specifically, deep learning is used to identify patterns within objective IT data, with subjective survey data serving as the ground truth. This allows for the validation of the customized theoretical values for customer satisfaction $S_{merit}$, $S_{merit}^R$, customer dissatisfaction $S_{demerit}$, consumer surplus $TS$, and the recipient price $p ^R$, as obtained from equations (65) and (67). Similarly, for blue-collar employees in both service and factory settings, measuring impact, employee satisfaction $S_{merit}^P$, employee dissatisfaction $S_{demerit}^{P}$, producer surplus $PS$, and the provider price $p^P$ for each process—based on a questionnaire or IT-based methods—enables the validation of the customized theoretical values obtained from equations (65) and (67). Furthermore, for white-collar employees in office or telework settings, similar validation can be performed by utilizing electroencephalographs, enterprise resource planning (ERP), and PC monitoring tools alongside the aforementioned metrics. In a laboratory setting, such validation can also be performed using a service field simulator (SFS), extended reality (XR), and magnetic resonance imaging (MRI) (Hara et al., 2023; Kawamura et al., 2025; Kurata et al., 2024; Masuda, 2022; Ogiso et al., 2023).

\section{Results}
When the repeat rate is equal to the rate of behavioral improvement due to nudges, the effect of nudges becomes greater because internalities are larger with environmental influences. The nudges considered in the framework of standard economics have the same structure as the nudges examined in this study. Gross social surplus is larger than in the case of standard economics (gross social surplus smaller than the value of the utility function in standard economics leads to high repeat rate). Reduced welfare due to bias can be expressed, and customized customer experiences can be designed. Considering IT within the context of business intelligence derived from service science and service engineering, this paper explains the connection between relevant variables and empirical validation.

\section{Discussion}
In general, nudges involving processes for making individual employees or customers perceive greater benefits and those using messages that equalize the weights of the loss process should lead to a higher rate of behavioral improvement. Examples include management of employees or customers which is strategic, reliable, and IT-driven and which does not depend on experience and intuition.

It should be possible to control biases through partial extraction of context (generalization of the environment or rules), though this is contingent on how it is done, and to control subjective well-being of both employees and customers by changing the value of intangible goods for customers. Therefore, heuristics represents a way to derive solutions that are roughly correct, resulting in high levels of happiness/satisfaction. 

The environment (rules) associated with intangible goods is analogous to a measurement device for a physical system. Therefore, in the absence of the environment (rules), the relevant intangible goods cannot be perceived (just like a physical system cannot be measured without a measurement device), which potentially prevent the establishment of intangible goods as a target of perception.

The environment (rules) can be formulated based on quantum information theory, thereby realizing measurement and control of the provider's and the recipient's perception using quantum computers, which could bring high economic benefits.

Because IT-enabled multimodal learning allows for the estimation of individual human emotions, integrating diverse IT data facilitates more varied and precise nudges. Ultimately, the ability to more accurately predict and influence individual decision-making has the potential to yield substantial economic outcomes. By repurposing this approach, the uncertain decision-making mechanisms of current “black-box” AI—which require massive datasets and incur high training costs—can be elucidated through quantum information theory and operationalized. This enables AI agents integrated with IT data to support or substitute for individual human decision-making more accurately and effectively, potentially driving significant economic benefits.

Analyzing the mechanism of nudges can potentially offer a window into the nature of human free will. In Bardeen–Cooper–Schrieffer (BCS) theory, cooling reduces the entropy of a system, causing the phase symmetry of electrons in a conductor to break as they transition into a superconducting state. An analogical counterpart of this concept in behavioral economics is that a nudge reduces the uncertainty (entropy) in decision-making, causing the symmetry of quality evaluation (the phase) in standard economics to break and giving rise to cognitive bias. In other words, it is possible to predict and influence individual decision-making by uncovering correlations between IT data and human-perceived product quality and human decision-making, and by managing the IT data correlated with the prior expectations that constitute that quality.

\section{Summary}
Based on quantum information theory, this study set up and analyzed an economic model of intangible goods, taking into account the relevant environment. Since the model is capable of explaining the relationship between bias and nudges, the approach based on quantum information theory can potentially be used for an economic model in the theory of design of intangible goods and a mathematical model for designing customer experiences. Considering IT within the context of business intelligence derived from service science and service engineering, this paper explained the connection between relevant variables and empirical validation.

\paragraph{Acknowledgments}
The author thanks Prof. T. Kurata of RIHSA at AIST, Prof. N. Nishino, and Assoc. Prof. T. Hara of the University of Tokyo for insightful discussions.


\section*{References}

\begin{description}

\item Allcott H and Taubinsky D (2015) Evaluating behaviorally motivated policy: Experimental evidence from the lightbulb market. American Economic Review 105(8): 2501–2538. Available at: 
\url{https://doi.org/10.1257/aer.20131564}.

\item Bernheim BD, Fradkin A and Popov I (2011) The welfare economics of default options: A theoretical and empirical analysis of 401 (k) plans. NBER Working Paper No. 17587. Cambridge, MA: National Bureau of Economic Research. Available at: \url{https://www.nber.org/system/files/working_papers/w17587/revisions/w17587.rev0.pdf}. (accessed 21 Sep. 2026).

\item Cheon T and Takahashi T (2010) Interference and inequality in quantum decision theory. Physics Letters A 375(2): 100–104. Available at: \url{https://doi.org/10.1016/j.physleta.2010.10.063}.

\item Ferrer-i-Carbonell A (2005) Income and well-being: An empirical analysis of the comparison income effect. Journal of Public Economics 89(5–6): 997‒1019. Available at: \url{https://doi.org/10.1016/j.jpubeco.2004.06.003}.

\item Fukuda M (2022) Proposal for reformulation of behavioral economic concepts centered on intangible goods by quantum information theory: Feasibility of mathematical design of customer experience. Journal of Behavioral Economics and Finance 15: 10–21. Available at: \url{https://doi.org/10.11167/jbef.15.10}. (in Japanese). 

\item Fukuda M (2023) Mathematical representation of bias and nudge centered on intangible goods by quantum information theory. Poster presented at the 10th International Congress on Industrial and Applied Mathematics (ICIAM 2023), Tokyo, Japan, 20–25 August. 

\item Fukuda M (2024) Mathematical representation of bias and nudges centered on intangible goods using quantum information theory. arXiv. Available at: \url{https://arxiv.org/abs/2411.08046}.

\item Grönroos C (1984) A service quality model and its marketing implications. European Journal of Marketing 18(4), 36‒44. Available at: \url{https://doi.org/10.1108/EUM0000000004784}.

\item Hara T, Hamano M, Ho BQ, Ota J, Yoshimoto Y and Arimitsu M (2023) Method for analyzing sequential services using EEG: Micro-meso analysis of emotional changes in real flight service. Physiology and Behavior 272: 114359. Available at: \url{https://doi.org/10.1016/j.physbeh.2023.114359}.

\item Jimenez-Gomez D (2018) Nudging and phishing: A theory of behavioral welfare economics. SSRN Electronic Journal. Available at:  \url{https://dx.doi.org/10.2139/ssrn.3248503}. (accessed 21 Sep. 2026). 

\item Kawamura H, Miura T, Maeda Y, Okada Y and Zempo K (2025) Framework for emotion recognition using cross-modal transformers with non-contact multimodal signals aiming clinical service support. IEEE Access 13: 99490‒99508. Available at: \url{https://doi.org/10.1109/ACCESS.2025.3573648}.
\item Kawasaki Y and Tadano K (2023) Design of behavior-inducing instruments utilizing nudges and behavioral economic methods: solving social issues by changing people’s daily behavior. Tokyo: Mitsubishi Research Institute Center for Advanced Technology. Available at:   \url{https://www.mri.co.jp/knowledge/column/dia6ou0000056z2g-att/3x_pdf_20230425.pdf}. (accessed 21 Sep. 2026). (In Japanese). 

\item Kurata T, Ogata K, Kanazawa S, Imamura Y, Sato A, Ogiso S, Kobayashi Y, Ichikari R, Nakae S, Tada M, Aoyama T, Shimizu H, Kuzuoka H, Nakamura T, Koshihara T, Kuroda M, Sorimachi H and Naruse F (2024) Project progress on XR-AI platform for tele-rehab and health guidance. In: 2024 IEEE Conference on Virtual Reality and 3D User Interfaces Abstracts and Workshops. Piscataway, NJ: IEEE, pp. 244‒251.
 Available at: \url{https://doi.org/10.1109/VRW62533.2024.00048}.

\item List JA and Samek AS (2015) The behavioralist as nutritionist: Leveraging behavioral economics to improve child food choice and consumption. Journal of Health Economics 39: 135–146. Available at: 
\url{https://doi.org/10.1016/j.jhealeco.2014.11.002}.

\item Masuda H and Matsumura K (2022) An empirical study on the impact of prior information provision methods on participants’ perceptions in tasting marketing. In: Stienmetz JL, Ferrer-Rosell B and Massimo D (eds) Information and Communication Technologies in Tourism 2022. ENTER 2022. Cham: Springer, pp. 455‒460. Available at:  \url{https://doi.org/10.1007/978-3-030-94751-4_42}.

\item Nagashima N and Nishio C (2013) The impact of evaluations of service process stages and the outcome on customer satisfaction: An empirical analysis benchmarked against prospect theory. Ryutsu Kenkyu 16(1): 35–59. Available at: \url{https://doi.org/10.5844/jsmd.16.1_35}. (in Japanese).

\item Nielsen MA and Chuang IL (2000) Quantum Computation and Quantum Information. Cambridge, UK: Cambridge University Press.

\item Ogiso S, Mori I, Miura T, Nakae S, Okuma T, Haga Y, Hatakeyama S, Kimura K, Kimura A and Kurata T (2023) Integration of BLE-based proximity detection with particle filter for day-long stability in workplaces. In: Proceedings of the IEEE/ION Position, Location, and Navigation Symposium (PLANS 2023). Piscataway, NJ: IEEE, pp. 1060–1065. Available at: \url{https://doi.org/10.1109/plans53410.2023.10140037}.

\item Shimomura Y, Hara T, Watanabe K, Sakao T, Arai T and Tomiyama T (2005) Proposal of the service engineering (1st report, service modeling technique for the service engineering). Transactions of the Japan Society of Mechanical Engineers Series C71(702): 315–322. Available at: \url{https://doi.org/10.1299/kikaic.71.669}. (in Japanese). 

\item Suwa Y (2015) Analysis of service value and price. Available at: \url{http://www.ipsj.or.jp/event/sj/sj2015/l2ist800000004ir-img/Session2.pdf}. (accessed 21 Sep. 2026) (in Japanese).

\item Takahashi T (2013) Rationality in quantum decision theory. Kagaku Tetsugaku 46(2): 17–30. Available at: 
\url{https://doi.org/10.4216/jpssj.46.17}. (in Japanese).

\item Thaler RH and Sustein CR (2008) Nudge: Improving Decisions About Health, Wealth and Happiness. New Haven, CT: Yale University Press.

\item Tversky A and Kahneman D (1992) Advances in prospect theory: Cumulative representation of uncertainty. Journal of Risk and Uncertainty 5: 297–323. Available at: 
\url{https://doi.org/10.1007/BF00122574}.

\item Vendrik M and Woltjer G (2006) Happiness and loss aversion: When social participation dominates comparison. IZA Discussion Paper No. 2218. Bonn: Institute for the Study of Labor. Available at:  \url{https://docs.iza.org/dp2218.pdf}.

\item von Neumann J and Morgenstern O (1944) Theory of Games and Economic Behavior. Princeton, NJ: Princeton University Press.

\item von Neumann J and Morgenstern O (1947) Theory of Games and Economic Behavior, Second Edition. Princeton, NJ: Princeton University Press.

\item Yukalov VI and Sornette D (2017) Quantum probabilities as behavioral probabilities. Entropy 19(3): 112. Available at: 
 \url{https://doi.org/10.3390/e19030112}.

\end{description}


\appendix
\section*{Appendix}

\setcounter{equation}{0}
\renewcommand{\theequation}{A\arabic{equation}}
The following is an outline of the value function for customer satisfaction in Fukuda (2022). It covers the equivalence between entropy and the utility function, the violation of the independence axiom, and the ratio of the value functions for gains and losses matching that in prospect theory. Furthermore, it discusses the fact that the value function mostly encompasses that of cumulative prospect theory and is consistent with econometric results.

According to Fukuda (2022), entropy has been proven equivalent to the utility function based on the minimum condition for the Kullback–Leibler divergence. Furthermore, as proven by Fukuda (2022), a phase shift in equations (42) and (43) alters their relative magnitude; this, in turn, changes the ordinal relationship between the value functions in (45) and reorders the preference relations. This may consequently result in a violation of the independence axiom that guarantees rationality in the framework of von Neumann and Morgenstern (1944, 1947). However, the independence axiom is not violated within the range of the value function in prospect theory.

Tversky and Kahneman (1992) conducted choice experiments by presenting various prospects to graduate students. By performing a nonlinear regression analysis on the results, they estimated a value function in cumulative prospect theory as follows.

\begin{equation} 
\nu(x) = 
  \begin{cases}
    x^{0.88}, & (x \ge 0), \\
    -2.25(-x)^{0.88}, & (x < 0).
  \end{cases}
\end{equation} 

For comparison, the mean of $\frac{S_{demerit}}{S_{merit}}$ for $0.2<p(\uparrow)<0.21$ and $0<x<10$ falls within the range $2.2<\overline{\frac{S_{demerit}}{S_{merit}}}<2.3$, which is approximately 2.25. In passing, the range of $p(\uparrow)$—the weight for emotional satisfaction that affects the repeat rate from equation (1)—is close to the values measured in the survey associated with Figure 2. 

These results imply that the value function for customer satisfaction obtained in this study mostly encompasses the value function in cumulative prospect theory. That is, for the sets $V(+)$, $V(-)$, $S (+)$, $S(-)$, the following holds.

\begin{equation} 
V(+)\subset S(+), \quad V(-)\subset S(-),
\end{equation} 

where

\begin{equation} 
V(+)=\{x|x\ge0,x^{0.88}\}, \quad V(-)=\{x|x\ge0,2.25x^{0.88}\},
\end{equation}

\begin{equation}
S(+)=\{x|x\ge0,S_{merit}\}, \quad S(-)=\{x|x\ge0,S_{demerit}\}.
\end{equation} 

According to Vendrik and Woltjer (2006), the standard specification used in the econometric study by Ferrer-i-Carbonell (2005) regarding life satisfaction and relative income is formulated as follows.

\begin{equation}
R=\alpha+\beta\log{Y}-\gamma\log{Y_r}+\delta X+\epsilon,
\end{equation} 

where $Y$ and $Y_r$ denote household income and the average household income of the social reference group, respectively; $\alpha$ is a parameter; $\beta$ and $\gamma$ are non-negative parameters; $\delta$ is a parameter vector; $\epsilon$  is the error term; and $X$ is a vector of control variables. Rearranging this yields the following.

\begin{equation}
R=\alpha+\beta\log{Y}-\gamma\log{Y}+\gamma\log{Y}-\gamma\log{Y_r}+\delta X+\epsilon \notag
\end{equation} 

\begin{equation}
=\alpha+(\beta-\gamma)\log{Y}+\gamma\log{\frac{Y}{Y_r}}+\delta X+\epsilon \notag
\end{equation} 

\begin{equation}
=\alpha+(\beta-\gamma)\log{Y}+\gamma\log{\left(1+\frac{Y}{Y_r}-1\right)}+\delta X+\epsilon \notag
\end{equation} 

\begin{equation}
=\alpha+(\beta-\gamma)\log{Y}+\gamma\log{\left(1+\frac{Y-Y_r}{Y_r}\right)}+\delta X+\epsilon.
\end{equation} 

The term of interest is extracted and is defined as subjective well-being $R^{'}$ as follows.

\begin{equation}
R^{'}=\gamma\log{\left(1+\frac{Y-Y_r}{Y_r}\right)}.
\end{equation} 

$R^{'}$ aligns with the value functions representing gains in (45), provided that $Y_r$ scales within $0<Y_r<1$, and that $\gamma=Y_r$ is chosen given that $\gamma$ is a non-negative parameter. Also, under the assumption of $Y-Y_{r}=x$, if $Y$  is in a typical household income range, $x$  falls within $0<x<10$  according to the scaling within $0<Y_r<1$, satisfying the evaluation range where the ratio of the value function for losses to that for gains is approximately $2$. Furthermore, applying the same mathematical form to define the value function for losses according to the sign of $x$ yields value functions equivalent to (45), as shown below.

\begin{equation}
R_{+}^{'}=\gamma_{+}\log{\left(1+\frac{x}{\gamma_+}\right)}, (x>0),
\end{equation} 

\begin{equation}
R_{-}^{'}=-\gamma_{-}\log{\left(1-\frac{x}{\gamma_-}\right)}, (x<0).
\end{equation} 

Since $\gamma_{+}$ and $\gamma_{-}$ are assumed to take on arbitrary positive values, these equations represent value functions in prospect theory in the case of $\gamma_{+}<\gamma_{-}$, while the generalized value function with $\gamma_{+}>\gamma_{-}$ represents subjective well-being.

\end{document}